\documentclass[11pt,a4paper]{article}
\usepackage[english]{babel}
\usepackage{amsmath,amsthm,amssymb,epsfig,latexsym}
\usepackage{color}
\usepackage{ulem}
\usepackage[numbers,square,sort&compress]{natbib}

\newcommand{\so}{\scriptscriptstyle \rm I}
\newcommand{\st}{\scriptscriptstyle \rm I\hspace{-1pt}I}

\newcommand{\bu}{\bar u}

\newcommand{\be}[1]{\begin{equation}\label{#1}}
\newcommand{\ba}[1]{\begin{multline}\label{#1}}
\newcommand{\ee}{\end{equation}}
\newcommand{\ea}{\end{multline}}

\newcommand{\qdet}{\mathop{\rm qdet}}
\newcommand{\sgn}{\mathop{\rm sgn}}

\newtheorem{thm}{Theorem}[section]
\newtheorem{prop}{Proposition}[section]
\newtheorem{lemma}{Lemma}[section]

\def\qed{\hfill\nobreak\hbox{$\square$}\par\medbreak}
 \makeatletter
 \@addtoreset{equation}{section}
 \makeatother
 
\newcommand{\bea}{\begin{eqnarray}}
\newcommand{\eea}{\end{eqnarray}}

\def\BB{{\mathbb{B}}}
\def\CC{{\mathbb{C}}}
\newcommand{\ZZ}{{\mathbb Z}}
\def\TT{{\mathbb{T}}}
\def\FF{{\rm F}}

\def\rvac{|0\rangle}
\def\lvac{\langle 0 |}

\def\EE{{\rm E}}
\def\FF{{\rm F}}
\def\hEE{\hat{\rm E}}
\def\hFF{\hat{\rm F}}
\def\hk{\hat{k}}
\def\TT{{T}}
\def\tEE{\tilde{\rm E}}
\def\tFF{\tilde{\rm F}}
\def\tTT{\tilde{T}}
\def\hTT{\widehat{T}}
\def\CF{\mathcal{F}}

\def\hCF{\hat{\mathcal{F}}}

\def\Ee{{\cal E}}
\def\hF{\hat{F}}
\def\hE{\hat{E}}

\begin{document}
\pagestyle{empty}
\setcounter{page}{0}

\begin{flushright}
LAPTH-036/18
\end{flushright}

\vspace{12pt}

\begin{center}
\begin{LARGE}
{\bf New symmetries of $\mathfrak{gl}(N)$-invariant Bethe vectors}
\end{LARGE}

\vspace{40pt}

\begin{large}
{A.~Liashyk${}^{a,b,c,d}$,
S.~Z.~Pakuliak${}^{e,f}$,\\ E.~Ragoucy${}^g$, N.~A.~Slavnov${}^h$\  \footnote{
a.liashyk@gmail.com, stanislav.pakuliak@jinr.ru, eric.ragoucy@lapth.cnrs.fr, nslavnov@mi-ras.ru}}
\end{large}

 \vspace{12mm}

${}^a$ {\it Bogolyubov  Institute for Theoretical Physics, NAS of Ukraine,  Kiev, Ukraine}

\vspace{2mm}

${}^b$ {\it National Research University Higher School of Economics, Faculty of Mathematics, Moscow, Russia}

\vspace{2mm}

${}^c$ {\it Skolkovo Institute of Science and Technology, Moscow, Russia}

\vspace{2mm}

${}^d$ {\it Institut Denis-Poisson, Université de Tours, Parc de Grandmont, 37200 Tours, France}

\vspace{2mm}

${}^e$ {\it Moscow Institute of Physics and Technology,  Dolgoprudny, Moscow reg., Russia}

\vspace{2mm}

${}^f$ {\it Laboratory of Theoretical Physics, JINR,  Dubna, Moscow reg., Russia}

\vspace{2mm}

${}^g$ {\it Laboratoire de Physique Th\'eorique LAPTh, CNRS and USMB,\\
BP 110, 74941 Annecy-le-Vieux Cedex, France}

\vspace{2mm}

${}^h$ {\it Steklov Mathematical Institute of Russian Academy of Sciences,\\ Moscow, Russia}

\end{center}

\vspace{4mm}

\begin{abstract}
We consider quantum integrable models solvable by the nested algebraic Bethe ansatz and possessing
$\mathfrak{gl}(N)$-invariant $R$-matrix. We study two types of Bethe vectors. The first type corresponds
to the original  monodromy matrix. The second type is associated to a
monodromy matrix closely related to the inverse of the monodromy matrix. We show that these two types of the Bethe vectors are
 identical up to normalization and reshuffling of the Bethe parameters. To prove this correspondence
we use the current approach.
This identity gives new combinatorial relations for the scalar products of the Bethe vectors. 
The 
$q$-deformed case, as well as the superalgebra case, are also  evoked in
the conclusion.
\end{abstract}

\newpage
\pagestyle{plain}

\section{Introduction}
The algebraic Bethe ansatz developed by the Leningrad school \cite{FadST79,FadT79,FadLH96} is a powerful method to investigate quantum integrable systems.
One can use this approach to find the spectra of quantum Hamiltonians.
Besides, this method can be used for calculating correlation functions of quantum integrable models \cite{BogIK93L,KitMT00,KitKMST12,GohKS04}.
In the framework of the algebraic Bethe ansatz this problem reduces to the calculating scalar products of Bethe vectors.

The notion of Bethe vector is one of the most important notions of the  algebraic Bethe ansatz. These vectors belong to the
physical space of states of the quantum model under consideration. They depend on a set of complex numbers called Bethe parameters.
Under certain constraints imposed on the Bethe parameters, the Bethe vector becomes an eigenvector of the quantum Hamiltonian.
In this case it is commonly called an {\it on-shell Bethe vector}. Otherwise, if the Bethe parameters are generic complex numbers,
the corresponding vector sometimes is called an {\it off-shell Bethe vector}.

In the  $\mathfrak{gl}(2)$ based model, the form of the Bethe vectors is quite simple \cite{FadST79,FadT79,FadLH96,BogIK93L}.
However, in the quantum integrable models with higher rank symmetry algebra, the construction of Bethe vectors becomes very
intricate. There are several ways to specify these vectors. A recursive procedure for constructing the off-shell Bethe vectors was
given in the papers \cite{KulRes81,KulRes82,KulRes83}. An explicit formula for these vectors (trace formula) containing tensor products of the
monodromy matrices and $R$-matrices was proposed in \cite{VT,VTcom,BeRa08}.
Another approach to
this problem, based on projections in the current algebra  was formulated in \cite{KhP-Kyoto,KhoPak05,KhoPakT07,FraKhoPR08}.
Explicit formulas for the Bethe vectors
in terms of the monodromy matrix entries acting on a reference  state were obtained in \cite{BelPRS12c,HutLPRS17a}.

In this paper we find a new symmetry of the Bethe vectors
in the models with $\mathfrak{gl}(N)$-invariant $R$-matrix. It is quite natural to expect that
the symmetries  of the monodromy matrix
should generate corresponding symmetries of the Bethe vectors \cite{KulRes83,VT,BelPRS12c,HutLPRS17a}.
In the present paper we consider a mapping of the  monodromy matrix $T$ to a new matrix $\widehat{T}$ closely related to the inverse monodromy matrix.
We study the properties of the Bethe vectors associated to the both these matrices. We show how these two types of Bethe vectors are related to each other.
As a direct application of this correspondence, we find new symmetries of the Bethe vector scalar products.

The paper is organized as follows.  We recall basic notions of the algebraic Bethe ansatz in section~\ref{Not}.
There we also give a notation used in the paper.
Section~\ref{BV} is devoted to the description of the properties of the Bethe vectors.
The main results of our paper are given in section~\ref{conn}, where we use an identification
of the Bethe vectors with certain combination of the generators of the Yangian double \cite{HutLPRS17a}
to prove the claimed symmetry of the Bethe vectors.
 In section~\ref{symm} we study symmetry properties of the scalar products of the Bethe vectors.
Several appendices gather technical details of the proofs.

\section{RTT-algebra and notation}\label{Not}
We consider quantum integrable models solvable by the algebraic Bethe ansatz and possessing $\mathfrak{gl}(N)$-invariant $R$-matrix
\begin{equation}\label{Rmat}
  R(u,v) = \mathbf{I}\otimes\mathbf{I} + g(u,v) \mathbf{P}, \quad g(u,v) = \frac{c}{u-v}.
\end{equation}
Here $\mathbf{I}=\sum_{i=1}^N\Ee_{ii}$ is the identity operator acting in the space $\mathbf{C}^N$, $\Ee_{ij}$ are $N\times N$ matrices
with the only nonzero entry
equal to $1$ at the intersection of the $i$-th row and $j$-th column,
$\mathbf{P}=\sum_{i,j=1}^N\Ee_{ij}\otimes\Ee_{ji} $ is the permutation operator acting in
$\mathbf{C}^N\otimes\mathbf{C}^N$, $c$ is a constant, and $u,v$ are arbitrary complex parameters called spectral parameters.

The key object of the algebraic Bethe ansatz is a monodromy matrix $T(u)$  with operator-valued
entries $T_{ij}(u)$ acting in a Hilbert space $\mathcal{H}$  (physical space of a quantum model). It satisfies an
$RTT$-algebra:
\begin{equation}\label{rtt}
  R(u,v) \left( T(u)\otimes\mathbf{I} \right) \left( \mathbf{I}\otimes T(v) \right) = \left( \mathbf{I}\otimes T(v) \right) \left( T(u)\otimes\mathbf{I} \right) R(u,v).
\end{equation}
Equation \eqref{rtt} yields the commutation relations  of the monodromy matrix  entries
\begin{equation}\label{rrt2}
  \left[ T_{ij}(u), T_{kl}(v) \right] = g(u,v) \left( T_{il}(u)T_{kj}(v) - T_{il}(v)T_{kj}(u) \right).
\end{equation}
Using (\ref{rtt}) it is easy to prove that
\begin{equation*}
  \left[ \mathcal{T} (u), \mathcal{T} (v) \right]=0,
\end{equation*}
where $\mathcal{T}(u)= \sum_{i} T_{ii}(u)$ is the transfer matrix. Thus, the transfer matrix is a generating function for the integrals
of motion of the model under consideration.

We assume the following dependence of the monodromy matrix elements $T_{ij}(u)$ on the parameter $u$
\begin{equation}\label{depen}
T_{ij}(u)=\delta_{ij}\mathbf{1}+\sum_{\ell\geq0} T_{ij}[\ell]u^{-\ell-1} ,
\end{equation}
where $\mathbf{1}$ and $T_{ij}[\ell]$ are respectively the unity and nontrivial operators acting in the Hilbert space $\mathcal{H}$.

\textsl{Remark.} In fact, for our purpose, the condition \eqref{depen} is optional. We impose this requirement on the asymptotics of $T(u)$ only in order to facilitate the presentation.
In quantum models of physical interest, the monodromy matrix may have a different asymptotic expansion, however, it can easily be reduced to the expansion \eqref{depen}.

We also assume that the space $\mathcal{H}$ has a pseudovacuum vector $\rvac$ (reference state) such that
\begin{equation}\label{tii}
\begin{aligned}
  & T_{ii}(u)\rvac  =  \lambda_{i}(u) \rvac, \\
  & T_{ij}(u)\rvac  =  0, \quad i>j,
\end{aligned}
\end{equation}
where $\lambda_i(u)$ are some functions  depending on the concrete quantum integrable model.
The action of $T_{ij}(u)$ with $i<j$ onto the pseudovacuum  is nontrivial. In the models of physical interest,
multiple  action of these operators onto $\rvac$ generates a basis in the space  $\mathcal{H}$.

Since the monodromy matrix is defined up to a common normalization scalar factor, it is convenient to deal with
the ratios:
\begin{equation}\label{ratios}
  \alpha_i(u) = \frac{\lambda_{i}(u)}{\lambda_{i+1}(u)}, \quad  i=1, \ldots, N-1  .
\end{equation}
We treat the functions $\alpha_i(u)$ as free functional parameters (generalized model) up to the restriction which follows from
\eqref{depen}.

Besides the original monodromy  matrix $T(u)$ we also can consider its inverse matrix. For this,
we first introduce the quantum determinant of the monodromy matrix $\qdet\bigl(T(u)\bigr)$ \cite{IzeK81,KulS82,MolNO96,Mol07} by
\begin{equation*}
  \qdet\bigl(T(u)\bigr) 	 = \sum_p \sgn(p) \; T_{1,{p(1)}}(u)\;T_{2,{p(2)}}(u - c)\;\ldots\; T_{N,{p(N)}}(u-(N-1)c).
\end{equation*}
Here the sum is taken over all permutations $p$ of the set $\{1,2,\ldots\,N\}$, $p(i)$ being the $i$-th element of the permutation $p$ of the set $\{1,2,\ldots\,N\}$. The quantum determinant  generates the center of the $RTT$-algebra
\begin{equation*}
  \left[\; \qdet\bigl(T(u)\bigr),\; T_{ij}(v) \; \right] = 0.
\end{equation*}
It is also easy to see that due to \eqref{tii}
\begin{equation*}
  \qdet\bigl(T(u)\bigr) \rvac = \lambda_{1}(u) \lambda_{2}(u-c) \ldots \lambda_{N}(u-(N-1)c) \rvac .
\end{equation*}

Similarly to the quantum determinant, we can introduce quantum minors of the size $m\times m$  ($1\le m<N$)
\begin{equation}\label{qminor}
 t^{a_1, a_2, \ldots, a_m}_{b_1, b_2, \ldots, b_m}(u) = \sum_p \sgn(p) \; T_{{a_1},b_{p(1)}}(u)\;T_{a_2, b_{p(2)}}(u - c)\;\ldots\; T_{a_m ,b_{p(m)}}(u-(m-1)c).
\end{equation}
Here the sum is taken over permutations of the set $\{1,2,\ldots\,m\}$, $p(i)$ being the $i$-th element of the permutation $p$ of the set $\{1,2,\ldots\,m\}$.

Now we can introduce the inverse monodromy matrix  $\tilde{T}(u)$
\begin{equation}\label{inver}
   \tilde{T}(u) T(u) =  \mathbf{I},
\end{equation}
where the entries $\tilde{T}_{ij}(u)$ are given by quantum minors divided by the quantum determinant
\begin{equation}\label{ttildef}
   \tilde{T}_{ij}(u) = (-1)^{i+j} t^{1\ldots\hat\jmath\ldots N}_{1\ldots\hat\imath\ldots N}(u-c)\ \mbox{qdet}(T(u))^{-1}.
\end{equation}
Here $\hat\imath$ and $\hat\jmath$ mean that the corresponding indices are omitted.

It is known \cite{Mol07} that the inverse monodromy matrix satisfies the $RTT$-relation with opposite sign of the constant $c$,
that is
\begin{equation*}
[ \tilde{T}_{ij}(u), \tilde{T}_{kl}(v) ] = g(v,u) \left( \tilde{T}_{il}(u)\tilde{T}_{kj}(v) - \tilde{T}_{il}(v)\tilde{T}_{kj}(u) \right).
\end{equation*}
Then, defining $\widehat{T}_{ij}(u)$ by
\begin{equation}\label{thatdef}
  \widehat{T}_{ij}(u) = \tilde{T}_{N+1-j, N+1-i}(u),
\end{equation}
we find that the elements $\widehat{T}_{ij}(u)$ satisfy commutation relations
\begin{equation*}
  [ \widehat{T}_{ij}(u), \widehat{T}_{kl}(v)] = g(u,v) \left( \widehat{T}_{il}(u)\widehat{T}_{kj}(v) - \widehat{T}_{il}(v)\widehat{T}_{kj}(u) \right).
\end{equation*}
Since these commutation relations coincide with \eqref{rrt2}, we conclude that $\widehat{T}(u)$ satisfies the
$RTT$-algebra \eqref{rtt} with the same $R$-matrix \eqref{Rmat}.

Thus, a mapping
\begin{equation}\label{mapp}
{T}_{ij}(u)\to \widehat{T}_{ij}(u)
\end{equation}
 is an automorphism of the $RTT$-algebra. The aim of this paper is
to investigate the symmetries of the off-shell Bethe vectors (see section \ref{BV}) related to this automorphism.

\subsection{Notation}
In this section we describe a notation that we use below.
First, we  introduce a special notation for the combination $1+g(u,v)$
\begin{equation}\label{f}
  f(u,v) = 1 + g(u,v) = \frac{u-v+c}{u-v}.
\end{equation}

Second,  we formulate a convention on the notation of sets of variables.
We denote them by bar:  $\bar t^i$, $\bar x^s$, and so on. Here
the superscripts refer to different sets.
Individual elements of the sets are denoted by subscripts:  $t^i_j$, $x^s_k$,  and so on.
Thus, for example, $\bar t=\{\bar t^1,\bar t^2\}$ means that the set $\bar t$ is the union of two sets
$\bar t^1$ and $\bar t^2$. At the same time, each of these two sets consists of the elements
$\bar t^s=\{t^s_1,t^s_2,\dots,t^s_{a_s}\}$, where $s=1,2$.

Notation $\bar t^i +\epsilon$ means that a constant $\epsilon$ is added to all the elements of the set $\bar t^i$.
Subsets of variables are denoted by roman indices: $\bar t^s_{\so}$, $\bar x^s_{\st}$, and so on.
In particular, we consider partitions of sets into subsets.
Then the notation $\{\bar t^s_{\so},\;\bar t^s_{\st}\}\vdash \bar t^s $ means that the
set $\bar t^s$ is divided into two disjoint subsets $\bar t^s_{\so}$ and $\bar t^s_{\st}$.
The order of the elements in each subset is not essential.

To make the  formulas more compact we use a shorthand notation for the products of functions depending on one or two variables.
Namely, if the function $f$ \eqref{f} depends on a set of variables (or two sets of variables), this means that one should take the product
over the corresponding set (or the double product over both sets).
For example,
\begin{equation}\label{SH-prod}
   f(u,\bar t^i)=\prod_{t^i_j\in\bar t^i} f(u, t^i_j), \qquad
   f(\bar t^s,\bar x^p)= \prod_{t^s_j\in\bar t^s}\prod_{x^p_k\in\bar x^p} f(t^s_j,x^p_k).
\end{equation}
We use the same prescription for the products of commuting operators, their vacuum eigenvalues $\lambda_i$  \eqref{tii}, and
the ratios of these eigenvalues $\alpha_i$ \eqref{ratios}
\begin{equation}\label{shpr} %shorthand product
  \lambda_i(\bar t^i)=\prod_{ t^i_j\in\bar t^i} \lambda_i( t^i_j),  \quad\quad
  \alpha_{i}(\bar t^i) = \prod_{ t^i_j\in\bar t^i} \alpha_{i}(t^i_{j}), \quad\quad
  T_{ij}(\bar t^s_{\so})= \prod_{ t^s_k\in\bar t^s_{\so}} T_{ij}( t^s_k).
\end{equation}
We will extend this convention for new functions that will appear later. Finally,
by definition, any product over the empty set is equal to $1$.
A double product is equal to $1$ if at least one of the sets is empty.

\section{Bethe vectors}\label{BV}

One of the main tasks of the algebraic Bethe ansatz is to find the eigenvectors of the transfer matrix, that usually
are called on-shell Bethe vectors. To do this, one should first construct off-shell Bethe vectors (or equivalently, Bethe vectors), that
belong to the Hilbert space $\mathcal{H}$. The latter are
special polynomials in $T_{ij}(u)$ with $i<j$ acting on $\rvac$.
In the simplest $\mathfrak{gl}(2)$ case the  Bethe vectors have the form $T_{12}(\bu)\rvac$, where
$\bu=\{u_1,\dots,u_n\}$, $n=0,1,\dots$. However, in the general $\mathfrak{gl}(N)$ case, the form of the
Bethe vectors is much more involved (see e.g. \cite{HutLPRS17a}).

In the $\mathfrak{gl}(N)$ based models, an off-shell Bethe vector $\BB(\bar t)$
depends on $N-1$ sets of complex numbers $\bar t=\{\bar{t}^1, \bar t^2, \ldots, \bar t^{N-1}\}$ called Bethe parameters.
The Bethe vector $\BB(\bar t)$ is symmetric over permutations of the Bethe parameters within each subset $\bar t^i$.
However, it is not symmetric with respect to rearrangements of subsets, and also for replacements $t^i_j \leftrightarrow t^k_l$.
If the Bethe parameters satisfy a special system of equations (Bethe equations), then off-shell Bethe vector becomes
an eigenstate of the transfer matrix. However, generically no constraint on the Bethe parameters $t^i_k$ are imposed.

Given a monodromy matrix $T(u)$, the different procedures\footnote{The known procedures are the nested algebraic Bethe ansatz \cite{KulRes81,KulRes82,KulRes83},
the trace formula \cite{VT,VTcom,BeRa08}, or the projection of currents \cite{KhP-Kyoto,KhoPak05,KhoPakT07,FraKhoPR08}.} to construct off-shell Bethe vectors provide, up to a global normalization factor, the same vectors, although several different explicit forms may exist  due to the commutation relations \eqref{rrt2}. Then, it remains to fix  unambiguously this normalization factor.
In this paper we use the same normalization
as in \cite{HutLPRS17b}. Namely, we have already mentioned  that a generic Bethe vector has the form of a polynomial
in $T_{ij}$ with $i<j$ applied to the pseudovacuum $|0\rangle$. Among all the terms of this polynomial, there is one monomial
that contains the operators $T_{ij}$ with $j-i=1$ only.
We call this term the {\it main term} and denote it by $\widetilde{\mathbb{B}}(\bar t)$. %Then
We fix the normalization of the Bethe vectors by fixing the numeric coefficient of the main term
\begin{equation}\label{mtb} %main term for B
 \widetilde{\mathbb{B}}(\bar t)=\frac{T_{N-1,N}(\bar t^{N-1})T_{N-2,N-1}(\bar t^{N-2}) \cdots
 T_{23}(\bar t^2) T_{12}(\bar t^1) |0\rangle}
				      {\prod_{i=1}^{N-1}\lambda_{i+1}(\bar t^{i})\prod_{i=1}^{N-2}  f(\bar t^{i+1},\bar t^i)}.
\end{equation}
Recall that we use here the shorthand notation \eqref{SH-prod}, \eqref{shpr} for the products of the operators
$T_{i,i+1}$, the vacuum eigenvalues $\lambda_{i+1}$, and the $f$-functions.

\subsection{Bethe vectors of the  matrix $\widehat{T}(u)$}\label{BVTI}

We have seen in the previous section that the matrix $\widehat{T}(u)$ satisfies the $RTT$-relation \eqref{rtt}.
Using the definition of $\widehat{T}_{ij}$ (see \eqref{ttildef}, \eqref{thatdef}, and \eqref{qminor}) one can find
the action of the operators $\widehat{T}_{ij}$ onto the pseudovacuum.
A straightforward calculation shows that
\begin{equation}\label{HTvac}
\begin{aligned}
  &\widehat{T}_{ij}(u)\rvac  = 0,  \qquad  i>j,\\
  &  \widehat{T}_{ii}(u)\rvac  =  \hat\lambda_{i}(u) \rvac,
  \end{aligned}
\end{equation}
where
\begin{equation}\label{Hlam}
  \hat\lambda_{i}(u)  = \frac{1}{\lambda_{N-i+1}(u-(N-i)c))}\prod_{\ell=1}^{N-i}\frac{\lambda_{\ell}(u-\ell c)}
  {\lambda_{\ell}(u-(\ell-1) c)} .
\end{equation}
It follows from \eqref{Hlam} that the ratios of the vacuum eigenvalues have the following form
\begin{equation}\label{hal}%hat alpha
  \hat\alpha_i(u) = \frac{\hat\lambda_{i}(u)}{\hat\lambda_{i+1}(u)} = \alpha_{N-i}(u-(N-i)c).
\end{equation}
Finally, the operators $\widehat{T}_{ij}$ with $i<j$ act on $\rvac$ as  creation operators.

Thus, we can construct  off-shell Bethe vectors $\hat\BB(\bar t)$ associated to the monodromy matrix $\widehat{T}(u)$. These
vectors are uniquely defined  provided their normalization is fixed. We do this as in \eqref{mtb}. Namely, the main
term $ \hat{\widetilde{\mathbb{B}}}(\bar t)$ of the off-shell Bethe vector
$\hat\BB(\bar t)$ reads
\begin{equation}
 \hat{\widetilde{\mathbb{B}}}(\bar t)= \frac{
\widehat{T}_{N-1,N}(\bar t^{N-1}) \widehat{T}_{N-2,N-1}(\bar t^{N-2})\cdots \widehat{T}_{23}(\bar t^2)\widehat{T}_{12}(\bar t^1)
 |0\rangle
}
				       {\prod_{i=1}^{N-1}\hat\lambda_{i+1}(\bar t^{i})\prod_{i=1}^{N-2}  f(\bar t^{i+1},\bar t^i)}.
\end{equation}
Here we have extended the shorthand notation \eqref{SH-prod}, \eqref{shpr} to the products of the operators
$\widehat{T}_{i,i+1}$ and the vacuum eigenvalues $\hat\lambda_{i+1}$.

The main result of this paper is a correspondence between $\BB(\bar t)$ and $\hat\BB(\bar t)$.

\section{Correspondence between two types of Bethe vectors}\label{conn}

In order to formulate the main result of this paper we introduce a mapping of the sets of  Bethe parameters:
\begin{equation}\label{map}
\mu(\bar t)\equiv \mu(\{\bar t^1,\bar t^2,\dots,\bar t^{N-1}\})=\{\bar t^{N-1}-c,\bar t^{N-2} - 2c,\dots,\bar t^{1}-(N-1)c\}.
\end{equation}
Thus, this mapping reorders the sets $\bar t^i$ and shifts every set $\bar t^i$ by $(i-N)c$.

\begin{thm}\label{Zidth}
The off-shell Bethe vectors $\BB$ and $\hat\BB$ of integrable models with $\mathfrak{gl}(N)$-invariant $R$-matrix
are related by
\begin{equation}\label{mid} %main identity
\hat\BB(\bar t)   =
  (-1)^{\# \bar t}\left( \prod_{s=1}^{N-2}  f(\bar t^{s+1},  \bar t^{s})\right)^{-1} \;\; \BB\bigl(\mu(\bar t)  \bigr).
\end{equation}
Here $\# \bar t$ is total cardinality of all the sets $\bar t^i$,
and according to \eqref{map}
\begin{equation}\label{shsets}
  \BB\bigl(\mu(\bar t)  \bigr) = \BB\bigl(\bar t^{N-1}-c,\bar t^{N-2} - 2c,\ldots,\bar t^{1}-(N-1)c\bigr).
\end{equation}
\end{thm}

 We prove this theorem using identification of the off-shell Bethe vectors with certain
combinations of the generating series of the Yangian double generators (see \cite{HutLPRS17a}).
The main tool of this approach relies on  the Gauss
coordinates of the monodromy matrix rather than considering  its matrix elements $T_{ij}(u)$.

\subsection{Gauss decomposition of the monodromy matrix}

The idea of using the Gauss decomposition of the monodromy matrix satisfying the $RTT$-relation \eqref{rtt}
goes back to the paper \cite{DF} where this decomposition was used to prove the isomorphism between
$R$-matrix and current realization of the quantum affine algebras.  Then the Gauss decomposition of the monodromy
was used in the series of papers
\cite{KhP-Kyoto,KhoPak05,KhoPakT07,FraKhoPR08}  to find closed and explicit formulas for the off-shell Bethe vectors. The Bethe vectors were expressed in terms
of the Gauss coordinates using a projection method
developed in those papers. In this section we find the relation between the Gauss coordinates
of the original $T(u)$ and the `transpose-inverse' monodromy $\widehat{T}(u)$. It will imply the
statement of  theorem~\ref{Zidth}.

As it was shown in the paper \cite{HutLPRS17a}, in order to obtain the off-shell Bethe vectors
in the form where the main term $\widetilde{\mathbb{B}}(\bar t)$ is  given by \eqref{mtb},
one has to use the following Gauss decomposition of the monodromy matrix $T(u)$ $($for $i<j)$:
\begin{align}\label{GF1}
\TT_{ij}(u)&=\FF_{j i}(u)k_{j}(u)+\sum_{j<\ell\leq N} \FF_{\ell i}(u)k_{\ell}(u)\EE_{j \ell}(u),\\
\label{GK1}
\TT_{ii}(u)&=k_{i}(u) +\sum_{i<\ell\leq N} \FF_{\ell i}(u)k_{\ell}(u)\EE_{i \ell}(u),\\
\label{GE1}
\TT_{ji}(u)&=k_{j}(u)\EE_{i j}(u)+\sum_{j<\ell\leq N}\FF_{\ell j}(u)k_{\ell}(u)\EE_{i \ell}(u).
\end{align}

These formulas are the result of product of three matrices
\begin{equation}\label{Gmat}
T(u)=\mathbf{F}(u)\cdot \mathbf{D}(u)\cdot \mathbf{E}(u)\,.
\end{equation}
In the above formula, $\mathbf{F}(u)$ is an upper-triangular matrix  with unities $\mathbf{1}$ on the diagonal,
$\mathbf{D}(u)=\mbox{diag}(k_1(u),k_2(u),\ldots,k_N(u))$ is a diagonal matrix,  and
$\mathbf{E}(u)$ is a
lower-triangular matrix  again with unities on the diagonal (see appendix~\ref{ApC}
for an example of these matrices in the case $N=3$).

It is clear from the reference state definition \eqref{tii} that the Gauss coordinates $\EE_{i j}(u)$
annihilate this state: $\EE_{i j}(u)\rvac=0$. The definition also implies that
it is a common eigenstate of the matrix  $\mathbf{D}(u)$ diagonal
 elements: $k_i(u)\rvac=\lambda_i(u)\rvac$ and that the
Gauss coordinates $\FF_{j i}(u)$ create non-trivial vectors in the space of states of the
quantum integrable models.

In order to describe the `transpose-inverse' monodromy matrix $\widehat{T}(u)$ in terms of the Gauss
coordinates $\FF_{j i}(u)$, $\EE_{i j}(u)$, $k_i(u)$ we have to invert the matrices
$\mathbf{F}(u)$, $\mathbf{D}(u)$ and $\mathbf{E}(u)$. The Gauss coordinates of
the inverse matrices
\begin{equation}\label{inm}
\begin{split}
\mathbf{F}(u)^{-1}&=\mathbf{I}+\textstyle{\sum_{i<j}}\Ee_{ij}\ \tFF_{j i}(u),\\
\mathbf{D}(u)^{-1}&=\mbox{diag}(k_1(u)^{-1},k_2(u)^{-1},\ldots,k_N(u)^{-1}),\\
\mathbf{E}(u)^{-1}&=\mathbf{I}+\textstyle{\sum_{i<j}}\Ee_{ji}\ \tEE_{i j}(u),
\end{split}
\end{equation}
are given by the following
\begin{lemma}
The Gauss coordinates $\tFF_{j i}(u)$ and $\tEE_{i j}(u)$, $1\leq i<j\leq N$ are
\begin{equation}\label{tFF}
\tFF_{j i}(u) =\sum_{\ell=0}^{j-i-1}(-)^{\ell+1}
\sum_{j>i_\ell>\cdots>i_1>i} \FF_{i_1,i}(u)  \FF_{i_2,i_1}(u)\cdots \FF_{i_\ell,i_{\ell-1}}(u)  \FF_{j,i_\ell}(u),
\end{equation}
\begin{equation}\label{tEE}
\tEE_{i j}(u) =\sum_{\ell=0}^{j-i-1}(-)^{\ell+1}
\sum_{j>i_\ell>\cdots>i_1>i} \EE_{i_\ell,j}(u)  \EE_{i_{\ell-1},i_\ell}(u)\cdots \EE_{i_1,i_2}(u)  \EE_{i,i_1}(u).
\end{equation}
\end{lemma}
{\it Proof} of this Lemma follows from a direct verification.\qed

According to the assumed dependence \eqref{depen}
of the monodromy matrix $T(u)$ on the spectral parameter $u$ we may conclude from the
formulas \eqref{GF1}--\eqref{GE1} that the Gauss coordinates $\FF_{j i}(u)$, $\EE_{i j}(u)$, $k_i(u)$
have the following dependence on the parameter $u$
\begin{equation}\label{dec}
\FF_{j i}(u)=\sum_{n\geq0}\FF_{j i}[n]u^{-n-1},\quad
\EE_{i j}(u)=\sum_{n\geq0}\EE_{i j}[n]u^{-n-1},\quad
k_i(u)=\mathbf{1}+\sum_{n\geq0}k_{i}[n]u^{-n-1}.
\end{equation}
The zero mode operators $\FF_{j i}[0]$, $\EE_{i j}[0]$ and $k_{i}[0]$ play an important role.
In particular,
according to the $RTT$ commutation relations \eqref{rtt} the Gauss coordinates with bigger difference
of the indices $j-i$ may be expressed as commutators of zero-mode operators and Gauss coordinates
with smaller difference $j-i$. In what follows we will need following
\begin{lemma}\label{zmcl}
The Gauss coordinates $\FF_{j i}(u)$, $\EE_{i j}(u)$ and $\tFF_{j i}(u)$, $\tEE_{i j}(u)$ can be written as
multiple commutators ($j>i$)
\begin{equation}\label{zm-comF}
\begin{split}
\FF_{j i}(u)&=c^{i+1-j}\Big[\Big[\cdots\Big[\Big[\FF_{j,j-1}(u),\FF_{j-1,j-2}[0]\Big],\FF_{j-2,j-3}[0]\Big],\cdots,
\FF_{i+2,i+1}[0]\Big],\FF_{i+1,i}[0]\Big],\\
\tFF_{j i}(u)&=-c^{i+1-j}\Big[\FF_{j,j-1}[0],\Big[\FF_{j-1,j-2}[0],\cdots,\Big[\FF_{i+3,i+2}[0],\Big[\FF_{i+2,i+1}[0],\FF_{i+1,i}(u)\Big]\Big]\cdots
\Big]\Big],
\end{split}
\end{equation}
and
\begin{equation}\label{zm-comE}
\begin{split}
\EE_{i j}(u)&=c^{i+1-j}\Big[\EE_{i,i+1}[0],\Big[\EE_{i+1,i+2}[0],\cdots,\Big[\EE_{j-3,j-2}[0],\Big[
\EE_{j-2,j-1}[0],\EE_{j-1,j}(u)\Big]\Big]\cdots\Big]\Big],\\
\tEE_{i j}(u)&=-c^{i+1-j}\Big[\Big[\cdots\Big[\Big[\EE_{i,i+1}(u),\EE_{i+1,i+2}[0]\Big],\EE_{i+2,i+3}[0]\Big],\cdots,\EE_{j-2,j-1}[0]\Big], \EE_{j-1,j}[0]\Big].
\end{split}
\end{equation}
\end{lemma}
{\it Proof}\,  is based on the $RTT$-relation for the monodromy matrix $T(u)$ and its inverse $\tilde{T}(u)$.
Details are given in appendix~\ref{ApB}.\qed

After applying  the transposition with respect to
the anti-diagonal to the inverse monodromy matrix $\tilde T(u)$, we obtain for the matrix $\widehat{T}(u)$ a Gauss decomposition $($for $i<j)$
\begin{align}\label{GF2}
\widehat{T}_{i j}(u)&=k_{N+1-j}(u)^{-1}\tFF_{N+1-i,N+1-j}(u)+\sum_{1\leq\ell< N+1-j}
\tEE_{\ell,N+1-j}(u)k_{\ell}(u)^{-1}\tFF_{N+1-i,\ell}(u),\\
\label{GK2}
\widehat{T}_{i i}(u)&=k_{N+1-i}(u)^{-1} +\sum_{1\leq\ell< N+1-i} \tEE_{\ell,N+1-i}(u)k_{\ell}(u)^{-1}\tFF_{N+1-i,\ell}(u),\\
\label{GE2}
\widehat{T}_{j i}(u)&=\tEE_{N+1-j,N+1-i}(u)k_{N+1-j}(u)^{-1}+\sum_{1\leq\ell< N+1-j}
 \tEE_{\ell,N+1-i}(u)k_{\ell}(u)^{-1} \tFF_{N+1-j,\ell}(u),
\end{align}
similar to the Gauss decomposition \eqref{GF1}--\eqref{GE1} of the original monodromy matrix $T(u)$.
The only crucial difference  is the ordering of the `new' Gauss coordinates in the formulas \eqref{GF2}--\eqref{GE2}.

We call a product of  the Gauss coordinates {\it normal ordered} if
all the coordinates $\FF_{j i}(u)$ are on the left of the product of all other Gauss coordinates and all
$\EE_{k l}(u)$ are on the right.
 This ordering is adapted to the action of the Gauss
coordinates onto reference state described above.

By construction, the expressions  \eqref{GF1}--\eqref{GE1} of the monodromy matrix
elements $T_{ij}(u)$ in terms of the Gauss coordinates
$\FF_{j i}(u)$, $\EE_{i j}(u)$, $i<j$ and $k_i(u)$, $i,j=1,\ldots,N$
 are written in the normal ordered form. However,
the formulas  \eqref{GF2}--\eqref{GE2}  for the inverse monodromy  matrix
 are not normal ordered. The normal ordering is given by the
  following
\begin{thm}\label{OrTh}
The normal ordered Gauss decomposition of the monodromy $\hTT(u)$ has  literally the same
form as in \eqref{GF1}--\eqref{GE1}  with the Gauss coordinates
$\FF_{j i}(u)$, $\EE_{i j}(u)$, $k_j(u)$ replaced by $\hFF_{j i}(u)$, $\hEE_{i j}(u)$, $\hk_j(u)$
where (for $i<j$)
\begin{align}\label{hFF}
\hFF_{j i}(u)&=\tFF_{N+1-i,N+1-j}(u-(N-j+1)c),\\
\label{hk}
\hk_{j}(u)&=\frac{1}{k_{N+1-j}(u-(N-j)c)} \prod_{\ell=1}^{N-j} \frac{k_\ell(u-\ell c)}{k_\ell(u-(\ell-1)c)} ,\\
\label{hEE}
\hEE_{i j}(u)&=\tEE_{N+1-j,N+1-i}(u-(N-j+1)c).
\end{align}
\end{thm}
{\it Proof}\,  is based on the presentation of the Gauss coordinates as multiple commutators. The shifts
of the indices in \eqref{hFF} and \eqref{hEE} can be seen from the formulas \eqref{GF2}
and \eqref{GE2}, while the shifts of the spectral parameters and transformation of the diagonal
generating series $k_j(u)\to \hk_j(u)$ follow from the commutation relations between
Gauss coordinates. They are gathered in appendix~\ref{ApC}. Note that formulas \eqref{hk} are in
accordance with the action of the diagonal matrix elements \eqref{HTvac} onto
the reference state $\rvac$. \qed

\subsection{Bethe vectors and currents}

This section is devoted to the proof of theorem~\ref{Zidth}.  We heavily use the results
of the paper \cite{HutLPRS17a} where the off-shell Bethe vectors were explicitly constructed
from the current generators of the  super-Yangian double
$DY(\mathfrak{gl}(m|n))$. In what follows we will use some results of this paper in the case
$m=N$, $n=0$.% and $R(u,v)$ is the $R$-matrix \eqref{Rmat} used in the commutation relations \eqref{rtt}.

The Yangian double associated with the algebra  $\mathfrak{gl}(N)$ is a Hopf algebra of a pair
of generating $N\times N$ matrices $T^\pm(u)$ satisfying the commutation relations
\begin{equation}\label{rtt-dy}
  R(u,v) \left( T^\kappa(u)\otimes\mathbf{I} \right) \left( \mathbf{I}\otimes T^\nu(v) \right) =
  \left( \mathbf{I}\otimes T^\nu(v) \right) \left( T^\kappa(u)\otimes\mathbf{I} \right) R(u,v) ,
\end{equation}
where $\kappa,\nu=\pm$. Being rewritten in terms  of the Gauss coordinates $\EE^\pm_{i j}(u)$,
$\FF^\pm_{j i}(u)$ and $k^\pm_i(u)$ \eqref{GF1}--\eqref{GE1} and generating series ({\it currents}) \cite{DF}
\begin{equation}\label{DF-iso1}
F_i({u})=\FF^{+}_{i+1,i}({u})-\FF^{-}_{i+1,i}({u})\,,\quad
E_i({u})=\EE^{+}_{i,i+1}({u})-\EE^{-}_{i,i+1}({u})\,,
\end{equation}
the commutation relations \eqref{rtt-dy} can be presented in the form
(so called `new' realization of the Yangian double)
\begin{equation}\label{tkiF}
\begin{split}
k^{\pm}_i(u) F_i(v) k^{\pm}_i(u)^{-1}&=
f(v,u)
\ F_i(v),\\
k^{\pm}_{i+1}(u)F_i(v)k^{\pm}_{i+1}(u)^{-1}&=
f(u,v)\ F_i(v),
\end{split}
\end{equation}
\begin{equation}\label{tkiE}
\begin{split}
k^{\pm}_i(u)^{-1}E_i(v)k^{\pm}_i(u)&=
f(v,u)\ E_i(v),\\
k^{\pm}_{i+1}(u)^{-1}E_i(v)k^{\pm}_{i+1}(u)&=
f(u,v)\ E_i(v),
\end{split}
\end{equation}
\begin{equation}\label{tFiFi}
f(u,v)\ F_i(u)F_i(v)=  f(v,u)\  F_i(v)F_i(u),
\end{equation}
\begin{equation}\label{tEiEi}
f(v,u)\ E_i(u) E_i(v)=  f(u,v)\  E_i(v) E_i(u),
\end{equation}
\begin{equation}\label{tFiFii}
(u-v-c)\ F_i(u)F_{i+1}(v)= (u-v)\ F_{i+1}(v)F_i(u),
\end{equation}
\begin{equation}\label{tEiEii}
(u-v)\ E_i(u)E_{i+1}(v)= (u-v-c)\  E_{i+1}(v)E_i(u),
\end{equation}
\begin{equation}\label{tEF}
\begin{split}
[E_i(u),F_j(v)]
&=c\ \delta_{i,j}\ \delta(u,v)\Big(k^+_{i}(u)\cdot k^+_{i+1}(u)^{-1}-k^-_{i}(v)\cdot k^-_{i+1}(v)^{-1}\Big),
\end{split}
\end{equation}
and the Serre relations for the currents $E_i(u)$ and $F_i(u)$. In \eqref{tEF} the symbol $\delta(u,v)$ means
the additive $\delta$-function given by the formal series
\begin{equation}\label{delta}
\delta(u,v)=\frac{1}{u}\sum_{\ell\in\ZZ}\frac{v^\ell}{u^\ell}\,.
\end{equation}

The Borel subalgebra in the Yangian double generated by matrix $T^+(u)$ is isomorphic
to the standard $\mathfrak{gl}(N)$ Yangian \cite{Mol07}. Then, we can identify the monodromy matrix $T(u)$ discussed in the previous sections
with the generating matrix $T^+(u)$. We also identify
the Gauss coordinates of these monodromy matrices
\begin{equation}\label{dec1}
\begin{split}
\FF^+_{j i}(u)&=\FF_{j i}(u)=\sum_{n\geq0}\FF_{j i}[n]u^{-n-1},\\
\EE^+_{i j}(u)&=\EE_{i j}(u)=\sum_{n\geq0}\EE_{i j}[n]u^{-n-1},\\
k^+_i(u)&=k_i(u)=\mathbf{1}+\sum_{n\geq0}k_{i}[n]u^{-n-1}.
\end{split}
\end{equation}

The currents $F_i(u)$, $k^+_j(u)$ and $E_i(u)$, $k^-_j(u)$ form the so-called dual Drinfeld Borel subalgebras with their own
Drinfeld coproduct properties. According to the general
theory of projections developed in \cite{EnrKP07}
one can define the projections $P_f^\pm$ and $P_e^\pm$ onto intersections of these current Borel subalgebras with
the standard Borel subalgebras formed by the Gauss coordinates
$\FF^+_{j i}(u)$, $\EE^+_{i j}(u)$, $k^+_j(u)$ and $\FF^-_{j i}(u)$, $\EE^-_{i j}(u)$, $k^-_j(u)$.

Due to the results of the papers \cite{KhP-Kyoto,HutLPRS17a} the  off-shell Bethe vectors can be identified
with the normalized projection of the product of the currents. In order to formulate this result we need to introduce
some notation.
For any scalar function $x(u,v)$ of two variables and any set $\bar u=\{u_1,\ldots,u_a\}$ we define the product\begin{equation}\label{tr-pr}
\Delta_x(\bu)=\prod_{i<j}\ x(u_j,u_i).
\end{equation}

Let $\CF_i(\bu)$, $i=1,\ldots,N-1$ be the ordered product of the currents
\begin{equation}\label{bv2}
\CF_i(\bu)=F_i(u_a)\cdot F_i(u_{a-1})\cdots F_i(u_2)\cdot F_i(u_1)\,.
\end{equation}
Note that  this product is not
symmetric with respect to permutation of the parameters $u_i$, as it follows from the commutation relation \eqref{tFiFi}.

One of the main result of the papers  \cite{KhP-Kyoto,HutLPRS17a}  is the
identification of the off-shell Bethe vectors with the projections of the product of the currents:% given  as follows
\begin{equation}\label{bv1}
\BB(\bar t)=\frac{\prod_{\ell=1}^{N-1}\Delta_f(\bar t^\ell)}{\prod_{\ell=1}^{N-2}f(\bar t^{\ell+1},\bar t^\ell)}
P_f^+\left(\CF_{N-1}(\bar t^{N-1})\CF_{N-2}(\bar t^{N-2})\cdots \CF_{2}(\bar t^{2})\CF_1(\bar t^1) \right)|0\rangle  .
\end{equation}
Observe that  the product
${\Delta}_f(\bar t^\ell)\CF_\ell(\bar t^\ell)$
is symmetric with respect to permutations within the set $\bar t^\ell$,  due to the commutation relations \eqref{tFiFi}.
As a result, the Bethe vector given by
equation \eqref{bv1} is symmetric with respect to the permutations of the Bethe parameters of the same type.

Mathematically rigorous definitions of the projections onto different type Borel subalgebras intersections can be found
in the paper \cite{EnrKP07}. They use the different Hopf structures associated with different type of Borel  subalgebras
in the Yangian double.
 However, one may understand the projection entering the equation \eqref{bv1} in a more simple way.
In order to calculate this projection one has to replace each current by the difference of the Gauss coordinates
\eqref{DF-iso1} and then use the commutation relations in the Yangian double \eqref{rtt-dy} between `positive' and `negative'
Gauss coordinates  sending all `negative' coordinates to the left and all `positive'
coordinates to the right. After such ordering  the action of the projection amounts to remove  all the terms
containing at least one `negative' Gauss coordinate on the left. Of course, practical implementation of this program
is rather heavy. Fortunately, there exist effective methods to perform this procedure \cite{KhP-Kyoto,HutLPRS17a}.

In this paper we are not going to describe the methods which allow to calculate the projection
in \eqref{bv1} and re-express the result of this calculation in terms of the original monodromy matrix element.
We refer the interested  reader to the paper  \cite{HutLPRS17a}. In order to prove the statement of theorem~\ref{Zidth}
we will need only the closed expression \eqref{bv1}.

The main trick in the calculation of the projection in \eqref{bv1} is the appearance of the so called
composed currents $F_{j i}(u)$, $i<j$ in the commutation relations of the currents
$F_{j,s+1}(u)$ and $F_{s i}(u)$ for $s=i+1,\ldots,j-1$. Then the rewriting of
the projection in \eqref{bv1} in terms of the monodromy matrix elements relies on the fact
that projections of the composed currents $P^+_f(F_{j i}(u))$ coincide with the Gauss
coordinates $\FF_{j i}(u)$ (see appendix~A of the paper  \cite{HutLPRS17a})
\begin{equation}\label{proF}
P^+_f(F_{j i}(u))=c^{j-i-1}\FF^+_{j i}(u)=c^{j-i-1}\FF_{j i}(u)\,.
\end{equation}

In order to prove the statement \eqref{mid} let us consider the rhs of this equality using the
expression \eqref{bv1}. We have
\begin{equation}\label{main-cal}
\begin{split}
&\frac{(-1)^{\# \bar t}} {\prod_{\ell=1}^{N-2}  f(\bar t^{\ell+1},  \bar t^{\ell})} \;\; \BB\bigl(\mu(\bar t)  \bigr)=\\
&\quad=\frac{(-1)^{\# \bar t}\prod_{\ell=1}^{N-1}\Delta_f(\bar t^{N-\ell}-{\ell} c)}
{\prod_{\ell=1}^{N-2}f(\bar t^{\ell+1},\bar t^\ell)f(\bar t^{N-\ell-1}-(\ell+1)c,\bar t^{N-\ell}-\ell c)}\\
&\qquad \times
P_f^+\left(\CF_{N-1}(\bar t^{1}-(N-1)c)\CF_{N-2}(\bar t^{2}-(N-2)c)\cdots
\CF_1(\bar t^{N-1}-c) \right)|0\rangle\\
&=\prod_{\ell=1}^{N-1}\Delta_f(\bar t^\ell)
P_f^+\left(\hCF_{1}(\bar t^{1})\hCF_{2}(\bar t^{2})\cdots
\hCF_{N-2}(\bar t^{N-2})\hCF_{N-1}(\bar t^{N-1}) \right)|0\rangle\\
&\quad=\frac{\prod_{\ell=1}^{N-1}\Delta_f(\bar t^\ell)} {\prod_{\ell=1}^{N-2}  f(\bar t^{\ell+1},  \bar t^{\ell})}
P_f^+\left(\hCF_{N-1}(\bar t^{N-1})\hCF_{N-2}(\bar t^{N-2})\cdots
\hCF_{2}(\bar t^{2})\hCF_{1}(\bar t^{1}) \right)|0\rangle.
\end{split}
\end{equation}
Here we have introduced the ordered product $\hCF_i(\bar t^i)$ of the shifted currents given by the product
\eqref{bv2} with the currents $F_i(u)$ replaced by the shifted currents $\hF_i(u)$
\begin{equation}\label{shif}
\hF_i(u)=-F_{N-i}(u-(N-i)c)\,.
\end{equation}
In \eqref{main-cal}, we also
used the identity $f(v,u)f(u-c,v)=1$ and the fact that the function  $f(u,v)$ is
translation invariant which implies $\Delta_f(\bu-\epsilon)=\Delta_f(\bu)$.
We also used the commutation relations
between currents  $\hF_i(u)$ and $\hF_{i+1}(v)$ which follow from \eqref{tFiFii}. The fact that one can use these
commutation relations under the action of the projection was proved in paper \cite{KhP-Kyoto}.

The assertion \eqref{mid} of theorem~\ref{Zidth} now follows from two lemmas.
\begin{lemma}\label{L43}
The mapping
\begin{equation}\label{map-cur}
\begin{split}
F_i(u)&\to \hF_i(u)=-F_{N-i}(u-(N-i)c),\quad i=1,\ldots,N-1 ,\\
E_i(u)&\to \hE_i(u)=-E_{N-i}(u-(N-i)c),\quad i=1,\ldots,N-1 ,\\
k^\pm_j(u)&\to \hk^\pm_j(u)=\frac{1}{k^\pm_{N+1-j}(u-(N-j)c)}
\prod_{\ell=1}^{N-j} \frac{k^\pm_\ell(u-\ell c)}{k^\pm_\ell(u-(\ell-1)c)},\quad j=1,\ldots,N
\end{split}
\end{equation}
is an automorphism of the Yangian double given by the commutation relations \eqref{tkiF}--\eqref{tEF}.
\end{lemma}
{\it Proof}\, is based on a direct verification. It is clear that the automorphism
\eqref{map-cur} is induced by the corresponding automorphism \eqref{mapp} of the $RTT$-algebra. \qed

\begin{lemma}\label{L44}
The projections of the composed currents $P^+_f(\hF_{j i}(u))$, $i<j$ which appear
in the commutation relations of the currents
$\hF_{j,s+1}(u)$ and $\hF_{s i}(u)$ for $s=i+1,\ldots,j-1$ coincide with the shifted Gauss coordinates
of the `transpose-inverse' monodromy matrix $\widehat{T}(u)$
\begin{equation}\label{prohF}
\begin{split}
P^+_f(\hF_{j i}(u))&=c^{j-i-1}\tFF^+_{N+1-i,N+1-j}(u-(N+1-j)c)\\
&=c^{j-i-1}\tFF_{N+1-i,N+1-j}(u-(N+1-j)c)
\end{split}
\end{equation}
given by the multiple commutators \eqref{zm-comF}.
\end{lemma}
{\it Proof}\, is given in appendix~\ref{ApB}. \qed

{\it Proof of theorem \ref{Zidth}.} As we can see from the equation \eqref{main-cal}
the Bethe vector $\hat\BB(\bar t)$ for the generalized quantum integrable models built from the
`transpose-inverse' monodromy matrix is given by the same formula as in \eqref{bv1} with currents $F_i(u)$
replaced by the currents $\hF_i(u)$. They satisfy
the same commutation relations \eqref{tkiF}--\eqref{tEF} with the currents $\hE_i(u)$ and $\hk^\pm_j(u)$ due to lemma~\ref{L43}.
Now using the statement of lemma~\ref{L44}
we can apply all the techniques developed in the papers \cite{KhP-Kyoto,HutLPRS17a}
and prove that $\hat\BB(\bar t)$ is the off-shell Bethe vector
 constructed from the monodromy matrix elements
$\hTT_{i j}(u)$ \eqref{thatdef}. Then, this proves the statement of theorem~\ref{Zidth}. \qed

\section{Symmetry of the highest coefficients}\label{symm}

As a direct application of
equation \eqref{mid}, we study  symmetry properties of the scalar products.
 For this, we should introduce dual Bethe vectors.

\subsection{Dual Bethe vectors}

 Dual Bethe vectors belong to the dual space
$\mathcal{H}^*$ and can be obtained by the successive action of $T_{j i}$ with $i<j$ from the right onto
a dual pseudovacuum $\langle0|\in \mathcal{H}^*$.
They also depend
on $N-1$ sets of complex numbers $\{\bar x^1,\bar x^2,\dots,\bar x^{N-1}\}$. Dual Bethe vectors
become dual eigenstates of the transfer matrix, if these parameters enjoy the system of Bethe equations.
For more details about these vectors, we refer the reader  to the works \cite{HutLPRS17a,HutLPRS17b}.

For the moment, it is important for us that the dual Bethe vectors can be obtained by a transposition
of ordinary Bethe vectors. Namely, a mapping $\psi\bigl(T_{ij}(u)\bigr)=T_{ji}(u)$ defines an anti-automorphism
of the $RTT$-algebra \cite{Mol07}:
\begin{equation}\label{anti}
   \psi(AB)=   \psi(B)   \psi(A).
\end{equation}
Here $A$ and $B$ are arbitrary products of the monodromy matrix entries $T_{ij}$. Extending this mapping to the
Bethe vectors by $\psi\bigl(\rvac\bigr)=\lvac$, one can prove that \cite{BelPRS12c,HutLPRS17a}
\begin{equation}\label{antiBV}
\CC(\bar x)= \psi\bigl( \BB(\bar x)\bigr),
\end{equation}
where $\CC(\bar x)$ is the dual Bethe vector.
Using this formula one can prove that the dual Bethe vectors also satisfy
a property similar to \eqref{mid}. Namely, let $\CC(\bar x)$ and $\hat\CC(\bar x)$  be dual Bethe vectors
respectively associated to the monodromy matrices $T(u)$ and $\widehat{T}(u)$. Then
\begin{equation}\label{midC} %main identity
\hat\CC(\bar x) =
  (-1)^{\# \bar x} \left(\prod_{s=1}^{N-2}  f(\bar x^{s+1},  \bar x^{s})\right)^{-1} \;\;   \CC\bigl(\mu(\bar x)\bigr).
\end{equation}
Here the notation is the same as in \eqref{mid}.

\subsection{Symmetries of the scalar products}

The scalar products of the Bethe vectors are defined as
\begin{equation}\label{SPdef}
S(\bar x|\bar t) =  \CC(\bar x) \BB(\bar t).
\end{equation}
The sets $\bar x$ and $\bar t$ are generic complex numbers such that $\#\bar x^i=\#\bar t^i$ for $i=1,\dots,N-1$. If the
latter condition does not hold, then the scalar product vanishes.

The scalar product  of generic Bethe vectors can be described by a {\it sum formula} \cite{HutLPRS17b}
\begin{equation}\label{sumfor}
  \CC(\bar x) \BB(\bar t) = \sum  W_{\text{\rm part}}(\bar x_{\so},\bar x_{\st}|\bar t_{\so},\bar t_{\st})
\prod_{k=1}^{N-1}  \alpha_k( \bar x^k_{\so}) \alpha_k(\bar t^k_{\st}).	
\end{equation}
Here all the sets of  Bethe parameters $\bar t^k$ and $\bar x^k$ are divided into two subsets $\{ \bar t^k_{\so} , \bar t^k_{\st} \}\vdash \bar t^k $
and $ \{ \bar x^k_{\so} , \bar x^k_{\st} \}\vdash \bar x^k$, such that $ \# \bar t^k_{\so} = \#\bar x^k_{\so}$.
The sum is taken over all possible partitions of this type.
The coefficients $W_{\text{\rm part}}$ are rational functions  completely determined by the $R$-matrix. They
do not depend on the ratios of the vacuum eigenvalues $\alpha_{k}$. Using the results of section~\ref{conn} we can
easily find symmetry properties of these coefficients.

\begin{prop}
For arbitrary partitions $\{ \bar t^k_{\so} , \bar t^k_{\st} \}\vdash \bar t^k $
and $ \{ \bar x^k_{\so} , \bar x^k_{\st} \}\vdash \bar x^k$, such that $ \# \bar t^k_{\so} = \#\bar x^k_{\so}$, the
corresponding coefficient $W_{\text{\rm part}}$ satisfies the following property:
\begin{equation}\label{WW1}
W_{\text{\rm part}}(\bar x_{\so},\bar x_{\st}|\bar t_{\so},\bar t_{\st})
\prod_{k=1}^{N-2}  f(\bar x^{k+1},  \bar x^{k})f(\bar t^{k+1},  \bar t^{k})
= W_{\text{\rm part}}\bigl(\mu(\bar x_{\so}),\mu(\bar x_{\st})|\mu(\bar t_{\so}),\mu(\bar t_{\st})\bigr),
\end{equation}
where $\mu(\bar x)$ is defined in \eqref{map}.
\end{prop}

{\sl Proof.} We compute  the scalar product in two different ways. First, performing
 in \eqref{sumfor} the replacements $\bar x^k\to \bar x^{N-k}-{k}c$ and $\bar t^k\to \bar t^{N-k}-{k}c$, we arrive at
\begin{multline}\label{sumforsh}
  \CC\bigl(\mu(\bar x)\bigr) \BB\bigl(\mu(\bar t)\bigr) = \sum
   W_{\text{\rm part}}\bigl(\mu(\bar x_{\so}),\mu(\bar x_{\st})|\mu(\bar t_{\so}),\mu(\bar t_{\st})\bigr)\\
\times \prod_{k=1}^{N-1}  \alpha_k( \bar x^{N-k}_{\so}-{k}c) \,\alpha_k(\bar t^{N-k}_{\st}-{k}c).	
\end{multline}
Due to \eqref{hal} we obtain
\begin{equation}\label{sumforsh1}
  \CC\bigl(\mu(\bar x)\bigr) \BB\bigl(\mu(\bar t)\bigr) = \sum
  W_{\text{\rm part}}\bigl(\mu(\bar x_{\so}),\mu(\bar x_{\st})|\mu(\bar t_{\so}),\mu(\bar t_{\st})\bigr)
\prod_{k=1}^{N-1} \hat \alpha_k( \bar x^{k}_{\so}) \hat\alpha_k(\bar t^{k}_{\st}).	
\end{equation}
Finally, using \eqref{mid} and \eqref{midC} we transform the lhs as follows:
\begin{equation}\label{sumforsh2}
  \hat\CC(\bar x) \hat\BB(\bar t)\prod_{k=1}^{N-2}  f(\bar x^{k+1},  \bar x^{k})f(\bar t^{k+1},  \bar t^{k}) = \sum
  W_{\text{\rm part}}\bigl(\mu(\bar x_{\so}),\mu(\bar x_{\st})|\mu(\bar t_{\so}),\mu(\bar t_{\st})\bigr)
\prod_{k=1}^{N-1} \hat \alpha_k( \bar x^{k}_{\so}) \hat\alpha_k(\bar t^{k}_{\st}).%\\
%
%\times  W_{\text{\rm part}}\bigl(\mu(\bar x_{\so}),\mu(\bar x_{\st})|\mu(\bar t_{\so}),\mu(\bar t_{\st})\bigr).
\end{equation}

On the other hand, the scalar product of the Bethe vectors $\hat\CC(\bar x)$ and $\hat\BB(\bar t)$ is given by the sum formula
\begin{equation}\label{Hsumfor}
  \hat\CC(\bar x) \hat\BB(\bar t) = \sum  W_{\text{\rm part}}(\bar x_{\so},\bar x_{\st}|\bar t_{\so},\bar t_{\st})
\prod_{k=1}^{N-1}  \hat\alpha_k( \bar x^k_{\so}) \hat\alpha_k(\bar t^k_{\st}).	
\end{equation}
Since the functions $\hat\alpha_i(u)$ are free functional parameters, the equations \eqref{sumforsh2} and \eqref{Hsumfor} can give the same result
if and only if the coefficients of every product of $\hat\alpha_i$ coincide. Thus, we arrive at \eqref{WW1}. \qed

In particular, we can consider a partition such that $\bar x_{\so}=\bar x$ and $\bar t_{\so}=\bar t$. Then respectively
$\bar x_{\st}=\bar t_{\st}=\emptyset$. The corresponding coefficient $W_{\text{\rm part}}$ is  called the \textbf{highest coefficient}.
We denote it by $Z(\bar x|\bar t)$:
\begin{equation}\label{HCdef}
Z(\bar x|\bar t)=W_{\text{\rm part}}(\bar x,\emptyset|\bar t,\emptyset).
\end{equation}
Then it follows immediately from \eqref{WW1} that
\begin{equation}\label{ZZ1}
Z\bigl(\mu(\bar x)|\mu(\bar t)\bigr) =
Z(\bar x|\bar t)
\prod_{k=1}^{N-2}  f(\bar x^{k+1},  \bar x^{k})f(\bar t^{k+1},  \bar t^{k})	.
\end{equation}

\section*{Conclusion}
In this paper we have found a new symmetry of Bethe vectors.
As we have mentioned, an off-shell Bethe vector is a polynomial in the monodromy matrix entries $T_{ij}$ applied to the pseudovacuum.
The new symmetry gives a description of the Bethe vector in terms of the entries of the monodromy matrix $\widehat{T}_{ij}$ \eqref{thatdef}.
%We have proved equivalence of those two representations.

%
In paper \cite{NL}, we have used already the symmetry of the Bethe vectors in the models with
$\mathfrak{gl}(3)$-invariant $R$-matrix. In that paper the equivalence of the two representations was
proved by the use of a recursion for the Bethe vectors. Generalization of this method to the case
of higher rank algebras is possible, but is technically very complex. Therefore,
our proof is based on the Gauss decomposition of the monodromy matrix and the underlying current algebra.
This approach was found to be very powerful in the study of the Bethe vectors for the models with high rank of symmetry \cite{HutLPRS17a}.

As a direct application of  the new symmetry, we proved  the identity for the highest coefficients of
the scalar product \eqref{ZZ1}. However, this is not the only possible application. The new representation
allows one to study the properties of combined operators that arise from the original monodromy
matrix $T_{ij}$ and from the  monodromy matrix $\widehat{T}_{ij}$. Recently this type of operators was considered in
\cite{GroLMS17}. There, in particular, it was conjectured that in $\mathfrak{gl}(3)$-invariant spin chains the operator
\begin{equation}\label{Bg-ND}
B^g(u)=T_{23}(u)\widehat{T}_{13}(u)-T_{13}(u)\widehat{T}_{12}(u)
\end{equation}
can be used for generating on-shell Bethe vectors. Our result allows us to obtain
explicit formulas for the action of $B^g(u)$ onto the Bethe vectors using known action formulas of the
operators $T_{ij}(u)$ \cite{BelPRS12c}. This allowed us to prove the conjecture of \cite{GroLMS17} and show that it is valid only for
 special (symmetric) representations of the Yangian \cite{NL}.

Concluding, we would like to mention that symmetries of the $RTT$-algebra, analogous to those considered in this paper, also exist for
the $RTT$-relations associated to the $U_q(\widehat{\mathfrak{gl}_n})$ algebras and  $\mathfrak{gl}(m|n)$ superalgebras.
As in the case discussed above, these symmetries generate new representations for the
Bethe vectors associated with the inverse monodromy matrix. In turn, these representations imply symmetries of scalar products,
in particular, symmetries of the highest coefficients. For the sake of completeness, we present the latter in the case of $U_q(\widehat{\mathfrak{gl}_n})$ and $\mathfrak{gl}(m|n)$ algebras.

For q-deformed algebra case $U_q(\widehat{\mathfrak{gl}_n})$, the highest coefficient $Z^q(\bar x|\bar t)$ was introduced in  \cite{HutLPRS18}.
Its symmetric property formally coincides with  \eqref{ZZ1}:
\begin{equation}\label{qZZ1}
Z^q\bigl(\mu(\bar x)|\mu(\bar t)\bigr) =
Z^q(\bar x|\bar t)
\prod_{k=1}^{n-2}  f^q(\bar x^{k+1},  \bar x^{k})f^q(\bar t^{k+1},  \bar t^{k})	,
\end{equation}
where
\begin{equation}
\mu(\bar t)=\{q^{-2}\, \bar t^{n-1} , q^{-4}\, \bar t^{n-2},\dots,q^{-2(n-1)}\, \bar t^{1}\}
\end{equation}
and
\begin{equation}
f^q(x,t)=\frac{q x - q^{-1} t}{x-t}.
\end{equation}
Relation \eqref{qZZ1} for the models described by $U_q(\widehat{\mathfrak{gl}_3})$ algebra was proven in \cite{PakRS13c}
via explicit representations for the highest coefficient.

For the superalgebra case $\mathfrak{gl}(m|n)$  (with $m,n>0$ and the grading $[i]=0$ for $i\le m$ and $[i]=1$ for $i>m$), the highest coefficient $Z^{n|m}(\bar x|\bar t)$ was introduced in \cite{HutLPRS17b}.
The relations between highest coefficients have slightly  more complex form:
\begin{multline}\label{ZZmn}
Z^{n|m}\bigl(\mu(\bar x)|\mu(\bar t)\bigr) =
(-1)^{\# \bar t^m}
\left. Z^{m|n}(\bar x|\bar t) \right|_{c \to -c} \\
\prod_{k=1}^{m-1}  f(\bar x^{k},  \bar x^{k+1})f(\bar t^{k},  \bar t^{k+1})	
 \prod_{k=m}^{n+m-2}  f(\bar x^{k+1},\bar x^{k}) f( \bar t^{k+1},\bar t^{k}),
\end{multline}

with
\begin{equation}
\mu(\bar t)=\{\bar t^{m+n-1}+(n-1)c,\bar t^{n+m-2} +(n-2) c,\dots,\bar t^{m+1} + c,\bar t^{m}, \bar t^{m-1} + c, \dots, \bar t^{1}+(m-1)c\}.
\end{equation}
 Note that equation \eqref{ZZmn} maps the highest coefficient of the scalar product in the $\mathfrak{gl}(m|n)$ superalgebra to that
of the scalar product in the $\mathfrak{gl}(n|m)$ superalgebra. The  map $c\to-c$  is specific to the superalgebra case  (see \cite{HutLPRS17b} for more details).

Let us stress once more that equations \eqref{qZZ1} and \eqref{ZZmn} are direct consequences of the symmetries of the Bethe vectors.
The latter can be proved exactly by the same method used in this paper.

\section*{Acknowledgments}
The work of A.L. has been funded by Russian Academic Excellence Project 5-100 and by Young Russian Mathematics award.
The work of S.P. was supported in part by the RFBR grant 16-01-00562-a. N.A.S. was supported by the
Russian Foundation RFBR-18-01-00273a.

\appendix

\section{Proof of lemmas~\ref{zmcl} and \ref{L44}}
\label{ApB}

We prove the statement of  lemma~\ref{zmcl}
using the commutation relations between Gauss coordinates. In order to obtain these
commutation relations from the $RTT$-relation \eqref{rrt2} we  use the approach of  paper
\cite{DF}. We also use the fact that we consider the generalized model,   and hence,
eigenvalues of the diagonal monodromy matrix elements are arbitrary functional parameters.
This means that after substitution of the Gauss decomposition formulas into commutation
relations \eqref{rrt2}, we obtain equations for all possible products of the currents
$k_i(u)k_j(v)$ after normal ordering of the Gauss coordinates according to the rules described before
theorem~\ref{OrTh}. In particular, we obtain
%kkk
\begin{equation}\label{ap3}
k_i(u)\FF_{i+i,i}(v)k_i(u)^{-1}=f(v,u)\FF_{i+i,i}(v)+g(u,v)\FF_{i+i,i}(u),
\end{equation}
\begin{equation}\label{ap4}
k_i(u)^{-1}\EE_{i,i+i}(v)k_i(u)=f(v,u)\EE_{i,i+i}(v)+g(u,v)\EE_{i,i+i}(u),
\end{equation}
\begin{equation}\label{ap5}
[\EE_{i,i+1}(v),\FF_{j+1,j}(u)]=\delta_{i,j}\, g(v,u)\left(k_i(u)k_{i+1}(u)^{-1}-k_i(v)k_{i+1}(v)^{-1}\right)\,,
\end{equation}
\begin{equation}\label{ap1}
\begin{split}
\FF_{j,j-1}(v)\FF_{j-1,i}(u)&=f(v,u)\FF_{j-1,i}(u)\FF_{j,j-1}(v)+\\
&+g(u,v)\Big(\FF_{j i}(v)-\FF_{j i}(u)+\FF_{j-1,i}(u)\FF_{j,j-1}(u)\Big)\,,
\end{split}
\end{equation}
\begin{equation}\label{ap2}
\begin{split}
\EE_{i,j-1}(u)\EE_{j-1,j}(v)&=f(v,u)\EE_{j-1,j}(v)\EE_{i,j-1}(u)+\\
&+g(u,v)\Big(\EE_{i j}(v)-\EE_{i j}(u)+\EE_{j-1,j}(u)\EE_{i,j-1}(u)\Big)\,.
\end{split}
\end{equation}

These equalities can be used to prove \eqref{zm-comF} and \eqref{zm-comE} respectively. Since both proofs are
identical we consider only \eqref{zm-comF}. Using the dependence of the Gauss coordinates on the spectral parameter
\eqref{dec1} we can send $u\to\infty$ {or} $v\to\infty$ and consider the  coefficients of the leading terms in \eqref{ap1}
at $u^{-1}$ {or} $v^{-1}$ respectively. We obtain
\begin{equation}\label{ap6}
\FF_{j i}(v)=c^{-1}[\FF_{j,j-1}(v),\FF_{{j-1},i}[0]]
\end{equation}
and
\begin{equation}\label{ap7}
\FF_{j i}(u)-\FF_{{j-1},i}(u)\FF_{j,j-1}(u)=c^{-1}[\FF_{j,j-1}[0],\FF_{{j-1},i}(u)].
\end{equation}
Now the first equation in \eqref{zm-comF} follows from a trivial induction of the relation \eqref{ap6}.
%\liashyk{Actually, no. One needs equation $\FF_{j,i}(v)=c^{-1}[\FF_{j,i+1}(v),\FF_{i+1,i}[0]]$ for it.}
%{\red\NNSS{\red Actually, yes. You assume the validness of \eqref{ap6} for $j$ and then prove that it is valid for $j+1$ using specialisation of \eqref{ap6} at $v\to\infty$}}.
By the induction over $j$, one can prove from \eqref{ap7} that following relation is valid
\begin{equation}\label{ap8}
\begin{split}
&c^{s-j}\Big[\FF_{j,j-1}[0],\Big[\FF_{j-1,j-2}[0],\cdots,\Big[\FF_{s+2,s+1}[0],\Big[\FF_{s+1,s}[0],\FF_{s i}(u)\Big]\Big]\cdots
\Big]\Big]=\\
&\qquad=\sum_{\ell=0}^{s-j}(-)^\ell\sum_{j>i_{\ell}>\cdots>i_1\geq s}
\FF_{i_1,i}(u)\FF_{i_2,i_1}(u)\cdots \FF_{i_\ell,i_{\ell-1}}(u)\FF_{j,i_\ell}(u),
\end{split}
\end{equation}
for any $s$ satisfying $i<s<j$. The second equality in \eqref{zm-comF} is a particular case of \eqref{ap8} at
$s=i+1$. This ends the proof of lemma \ref{zmcl}.\qed

In order to prove the statement of lemma~\ref{L44} we use the results of the appendix~A of  paper
\cite{HutLPRS17a}. We consider  the shifted currents $\hF_i(u)$ \eqref{shif} and the corresponding composed  currents
$\hF_{j i}(u)$ defined in this appendix by the formulas (A.3) and (A.7).
 These {composed} currents
satisfy a relation identical to (A.17) {in the same appendix of \cite{HutLPRS17a}}, which implies
\begin{equation}\label{ap9}
P^+_f\Big(\hF_{j i}(u)\Big)=\Big[P^+_f\Big(\hF_{j,i+1}(u)\Big),\hF_i[0]\Big].
\end{equation}
The commutativity between the projections and commutation relations with zero modes was
proved in appendix~B of \cite{HutLPRS17a}. Now the chain of equations ($i'=N+1-j$ and $j'=N+1-i$)
\begin{equation}\label{ap10}
\begin{split}
&P^+_f\Big(\hF_{j i}(u)\Big)=\Big[\Big[\cdots\Big[\Big[P^+_f\Big(\hF_{j-1}(u)\Big),\hF_{j-2}[0]\Big],\hF_{j-3}[0]\Big],\cdots,\hF_{i+1}[0]\Big],\hF_i[0]\Big]\\
&\quad=-\Big[F_{N-i}[0],\Big[F_{N-i-1}[0],\cdots,\Big[F_{N+2-j}[0],P^+_f\Big(F_{N+1-j}(u-(N+1-j)c)\Big)\Big]\cdots\Big]\Big]\\
&\quad=-\Big[\FF_{j',j'-1}[0],\Big[\FF_{j'-1,j'-2}[0],\cdots,\Big[\FF_{i'+2,i'+1}[0],\FF_{i'+1,i'}(u-i'c)\Big]\cdots\Big]\Big]\\
&\quad=c^{j'-i'-1}\tFF_{j' i'}(u-i'c)=c^{j-i-1}\tFF_{N+1-i,N+1-j}(u-(N+1-j)c)
\end{split}
\end{equation}
proves relation \eqref{prohF}. This ends the proof of lemma~\ref{L44}.\qed

\section{Gauss coordinates and proof of theorem~\ref{OrTh} }
\label{ApC}

Before starting the proof of theorem~\ref{OrTh} we provide explicit formulas for the Gauss decomposition
 used in this paper in the simplest nontrivial case $N=3$. The monodromy matrix reads
\begin{equation}\label{ca1}
\begin{split}
\TT(u)&=\left(\begin{array}{ccc}
k_1+\FF_{21}k_2\EE_{12}+\FF_{31}k_3\EE_{13}& \FF_{21}k_2+\FF_{31}k_3\EE_{23}&\FF_{31}k_3\\
k_2\EE_{12}+\FF_{32}k_3\EE_{13}&k_2+\FF_{32}k_3\EE_{23}& \FF_{32} k_3\\
k_3\EE_{13}& k_3\EE_{23}&k_3
\end{array} \right)=\\
&=\left(\begin{array}{ccc}1&\FF_{21}&\FF_{31}\\0&1&\FF_{32}\\0&0&1 \end{array}\right)
\left(\begin{array}{ccc}k_1&0&0\\0&k_2&0\\0&0&k_3 \end{array}\right)
\left(\begin{array}{ccc}1&0&0\\\EE_{12}&1&0\\ \EE_{13}&\EE_{23}&1 \end{array}\right).
\end{split}
\end{equation}
For brevity, we omitted in \eqref{ca1} the dependence on the
spectral parameter $u$ for all Gauss coordinates $\EE_{ij}(u)$, $\FF_{ji}(u)$, and $k_i(u)$.

The Gauss decomposition \eqref{ca1} allows one to find easily the inverse monodromy matrix
\begin{equation}\label{ca2}
\begin{split}
\tTT(u)=\TT(u)^{-1}&=
\left(\begin{array}{ccc}1&0&0\\ \tEE_{12}&1&0\\ \tEE_{13}&\tEE_{23}&1 \end{array}\right)
\left(\begin{array}{ccc}k_1^{-1}&0&0\\0&k_2^{-1}&0\\0&0&k_3^{-1} \end{array}\right)
\left(\begin{array}{ccc}1&\tFF_{21}&\tFF_{31}\\0&1&\tFF_{32}\\0&0&1 \end{array}\right)=\\
&=
\left(\begin{array}{ccc}
k_1^{-1}& k_1^{-1}\tFF_{21}&k_1^{-1}\tFF_{31}\\
\tEE_{12}k_1^{-1}&k_2^{-1}+\tEE_{12}k_1^{-1}\tFF_{21}&k_2^{-1}\tFF_{32}+\tEE_{12}k_1^{-1}\tFF_{31} \\
\tEE_{13}k_1^{-1}& \tEE_{23}k_2^{-1}+\tEE_{13}k_1^{-1}\tFF_{21}
&k_3^{-1}+\tEE_{23}k_2^{-1}\tFF_{32}+\tEE_{13}k_1^{-1}\tFF_{31}
\end{array} \right)
\end{split}
\end{equation}
where
\begin{equation}\label{ca3}
\begin{split}
\tFF_{12}(u)=-\FF_{12}(u),\quad \tFF_{23}(u)=-\FF_{23}(u),\quad
\tFF_{31}(u)=-\FF_{31}(u)+\FF_{21}(u)\FF_{32}(u),\\
\tEE_{12}(u)=-\EE_{12}(u),\quad \tEE_{23}(u)=-\EE_{23}(u),\quad
\tEE_{13}(u)=-\EE_{13}(u)+\EE_{23}(u)\EE_{12}(u).
\end{split}
\end{equation}

Now the monodromy matrix $\hTT(u)$ given by the relation \eqref{thatdef} has the following
structure:
\begin{equation}\label{ca4}
\hTT(u)=
\left(\begin{array}{ccc}
k_3^{-1}+\tEE_{23}k_2^{-1}\tFF_{32}+\tEE_{13}k_1^{-1}\tFF_{31}&k_2^{-1}\tFF_{32}+\tEE_{12}k_1^{-1}\tFF_{31} &k_1^{-1}\tFF_{31}\\
\tEE_{23}k_2^{-1}+\tEE_{13}k_1^{-1}\tFF_{21}&k_2^{-1}+\tEE_{12}k_1^{-1}\tFF_{21}&k_1^{-1}\tFF_{21} \\
\tEE_{13}k_1^{-1}&\tEE_{12}k_1^{-1}
&k_1^{-1}
\end{array} \right).
\end{equation}
It is similar to the structure of  the original monodromy matrix $\TT(u)$ \eqref{ca1}.

We prove theorem~\ref{OrTh} by induction starting from the right-lower corner of the
monodromy matrix $\hTT(u)$. Due to the formulas \eqref{GF2}--\eqref{GE2}
the matrix elements from the right-lower corner $\hTT_{N N}(u)$, $\hTT_{N-1,N}(u)$ and
$\hTT_{N,N-1}(u)$ have following form:
\begin{equation}\label{b1}
\hTT_{N N}(u)=k_1(u)^{-1},\quad \hTT_{N-1,N}(u)=k_1(u)^{-1}\tFF_{2 1}(u),\quad  \hTT_{N,N-1}(u)=\tEE_{1 2}(u)k_1(u)^{-1}\,.
\end{equation}
In order to normal order these matrix elements we can use the commutation relations \eqref{ap3} and \eqref{ap4}
specialised to $i=1$ and $v=u-c$. This yields
\begin{equation}\label{b4}
\hTT_{N-1,N}(u)=k_1(u)^{-1}\FF_{2 1}(u)=\FF_{2 1}(u-c)k_1(u)^{-1},
\end{equation}
\begin{equation}\label{b5}
 \hTT_{N,N-1}(u)=\EE_{1 2}({u})k_1(u)^{-1}=k_1(u)^{-1}\EE_{1 2}(u-c),
\end{equation}
and proves formulas \eqref{hFF} and \eqref{hEE} in the particular case $i=N-1$ and $j=N$.
Now using \eqref{ap5}  at $i=1$ and \eqref{b4}, \eqref{b5} we can normal order the monodromy matrix element
\begin{equation*}
\hTT_{N-1,N-1}(v)=k_2(v)^{-1}+\tEE_{1 2}(v)k_1(v)^{-1}\tFF_{2 1}(v)
\end{equation*}
to obtain
\begin{equation*}
\begin{split}
&\EE_{1 2}(v)k_1(v)^{-1}\FF_{2 1}(v)= \EE_{1 2}(v)\FF_{2 1}(v-c)k_1(v)^{-1}=\\
&=\FF_{2 1}(v-c)k_1(v)^{-1}\EE_{1 2}(v-c)+\frac{k_1(v-c)}{k_2(v-c)k_1(v)}-k_2(v)^{-1}\,.
\end{split}
\end{equation*}
As a result, the element $\hTT_{N-1,N-1}(v)$ in the normal ordered form is equal to
\begin{equation}\label{b7}
\hTT_{N-1,N-1}(v)=\frac{k_1(v-c)}{k_2(v-c)k_1(v)}+\tFF_{2 1}(v-c)k_1(v)^{-1}\tEE_{1 2}(v-c),
\end{equation}
thus proving \eqref{hk} for $j=N-1$.

Formulas \eqref{b4}, \eqref{b5}, and \eqref{b7} are the base of the induction. Let us assume that the statement
of theorem~\ref{OrTh} is valid for $\ell\leq i<j\leq N$ in \eqref{hFF},\eqref{hEE} and
for $\ell\leq j\leq N$ in \eqref{hk}. By exploring the commutation relations between the Gauss
coordinates and lemma~\ref{zmcl} we will prove that these formulas are valid for $\ell\to\ell-1$.

Let us consider the commutation relation \eqref{rrt2} for the monodromy
matrix elements $\hTT_{i j}(u)$ at the values of indices $(i,j,k,l)\to(\ell-1,j,j,j)$
and send $u\to \infty$. Then the coefficient  of $u^{-1}$ gives (for $j=\ell,\ldots,N$)
\begin{equation}\label{b8}
\hTT_{\ell-1,j}(u)=c^{-1}\Big[\hTT_{j j}(u),\hTT_{\ell-1,j}[0]\Big].
\end{equation}
The zero mode of the monodromy matrix element $\hTT_{\ell-1,j}[0]$ can be obtained from the relation
\eqref{GF2} and is equal to  %\liashyk{one should admit that $ l-1 < j$}{\red \NNSS{\red It is written before \eqref{b8}.}}
\begin{equation}\label{b9}
\hTT_{\ell-1,j}[0]=\tFF_{\ell'+1,j'}[0] ,
\end{equation}
where here and below the prime on the index $j$  mean  $j'=N+1-j$ for any index $j$.

According to the induction assumption, the monodromy matrix elements  $\hTT_{j j}$
($j=\ell,\ldots,N$) have the normal ordered form
\begin{equation}
\label{GK3}
\hTT_{j j}(u)=\hk_{j}(u) +\sum_{j<s\leq N} \hFF_{s j}(u)\hk_{s}(u)\hEE_{j s}(u),
\end{equation}
where the Gauss coordinates $\hFF_{s j}(u)$, $\hk_{s}(u)$, and $\hEE_{j s}(u)$  respectively are given
by equations \eqref{hFF}, \eqref{hk} and \eqref{hEE}.
One can prove from the commutation relations between the Gauss coordinates
that the zero mode $\hTT_{\ell-1,j}[0]$ \eqref{b9} commutes with
$\hk_{i}(u)$ and $\hFF_{s i}(u)$ $\forall i$, except for  $\hk_{j}(u)$ and $\hFF_{s j}(u)=\tFF_{j' s'}(u-s'c)$. These commutation relations are
\begin{equation}\label{b10}
c^{-1}\Big[\hk_j(u),\tFF_{\ell'+1,j'}[0]\Big]=\hFF_{j,\ell-1}(u)\hk_j(u)
\end{equation}
and
\begin{equation}\label{b11}
c^{-1}\Big[\hFF_{s j}(u),\tFF_{\ell'+1,j'}[0]\Big]=\tFF_{\ell'+1,s'}(u-s'c)=\hFF_{s,\ell-1}(u).
\end{equation}
To obtain \eqref{b10} we used the second relation in \eqref{zm-comF}, the commutation
relation
\begin{equation*}
[k_i(v)^{-1},\FF_{i+1,i}[0]]=c\,k_i(v)^{-1}\FF_{i+1,i}(v)=c\,\FF_{i+1,i}(v-c)k_i(v)^{-1},
\end{equation*}
which follows from \eqref{ap3}, and the commutativity $[k_i(v),\FF_{j+1,j}(u)]=0$ for $j>i$.
Equalities \eqref{b10} and \eqref{b11} imply that the rhs of \eqref{b8} is (for $j=\ell,\ldots,N$)
\begin{equation}\label{GF3}
\hTT_{\ell-1,j}(u)=\hFF_{j,\ell-1}(u)\hk_{j}(u)+\sum_{j<s\leq N} \hFF_{s,\ell-1}(u)\hk_{s}(u)\hEE_{j s}(u).
\end{equation}

Similarly we can prove that the commutation relations between the Gauss coordinates yield
\begin{equation}\label{GE3}
\hTT_{j,\ell-1}(u)=\hk_{j}(u)\hEE_{\ell-1,j}(u)+\sum_{j<s\leq N} \hFF_{s j}(u)\hk_{s}(u)\hEE_{\ell-1,s}(u) ,
\end{equation}
where the Gauss coordinates $\hFF_{s,\ell-1}(u)$ and $\hEE_{\ell-1,s}(u)$ are given by
\eqref{hFF} and \eqref{hEE} for $s=\ell,\ldots,N$.

To finish the proof of the theorem we have to prove that  the Gauss coordinates
$\hFF_{s,\ell-1}(u)$, $\hEE_{\ell-1,s}(u)$ and $\hk_s(u)$ given by the equalities
\eqref{hFF}--\eqref{hEE} for $s=\ell,\ldots,N$ will imply the same structure of the
Gauss coordinate $\hk_{\ell-1}(u)$.

To do this we can use  again  the commutation relations \eqref{rrt2} for $(i,j,k,l)\to(\ell-1,\ell,\ell,\ell-1)$
to obtain in the limit $v\to\infty$
\begin{equation}\label{b12}
\Big[\hTT_{\ell-1,\ell}(u),\hTT_{\ell,\ell-1}[0]\Big]=c\left(\hTT_{\ell \ell}(u)-\hTT_{\ell-1,\ell-1}(u)\right)\,,
\end{equation}
where the zero mode operator $\hTT_{\ell,\ell-1}[0]$ can be deduced from \eqref{GE2}
\begin{equation*}
\hTT_{\ell,\ell-1}[0]=-\EE_{N+1-\ell,N+2-\ell}[0]=-\EE_{\ell',\ell'+1}[0]\,.
\end{equation*}
Now the proof of \eqref{hk} for $\hk_{\ell-1}(u)$ follows from the inductive assumption
\eqref{GK3} and the commutation relations
\begin{equation*}
\Big[\EE_{j,j+1}[0],\FF_{j+1,j}(u)\Big]=c(k_j(u)k_{j+1}(u)^{-1}-1)\,,
\end{equation*}
\begin{equation*}
\Big[\EE_{j,j+1}[0],\FF_{j+1,j}[0]\Big]=c(k_j[0]-k_{j+1}[0])\,,
\end{equation*}
\begin{equation*}
\Big[\EE_{j,j+1}[0],k_{j}(u)^{-1}\Big]=ck_{j}(u)^{-1}\EE_{j,j+1}(u-c)\,,
\end{equation*}
\begin{equation*}
\Big[\EE_{j,j+1}[0],\tFF_{j+1,\ell}(u)\Big]=c\tFF_{j \ell}(u)\,,
\end{equation*}
and
\begin{equation*}
\Big[\EE_{j,j+1}[0],\tEE_{\ell j}(u)\Big]=-c\tEE_{\ell,j+1}(u)\,.
\end{equation*}
This finishes the proof of  theorem~\ref{OrTh}.\qed

\end{document}